\documentclass[floatfix,aps,prapplied,amsmath,amssymb,reprint]{revtex4-2}

\usepackage[utf8]{inputenc}
\usepackage[T1]{fontenc}
\usepackage{amsfonts}
\usepackage{graphicx}
\usepackage{siunitx}
\usepackage{color}
\usepackage{float}
\usepackage{physics} 
\usepackage{multirow} 
\usepackage[colorlinks=true,citecolor=blue,urlcolor=red]{hyperref}

\begin{document}

\title{
Accelerating the heat diffusion: Fast thermal relaxation of a microcantilever}
\author{Basile Pottier}
\affiliation{Univ Lyon, ENS de Lyon, CNRS, Laboratoire de Physique, F-69342 Lyon, France}

\author{Carlos A. Plata}
\affiliation{Fisica Te\'orica, Universidad de Sevilla, Apartado de Correos 1065, E-41080 Sevilla, Spain}

\author{Emmanuel Trizac}
\affiliation{Universit\'e Paris-Saclay, CNRS, LPTMS, F-91405 Orsay, France}

\author{David Gu\'ery-Odelin}
\affiliation{Université de Toulouse, CNRS, IRSAMC, Laboratoire de Collisions Agrégats Réactivité, F-31062 Toulouse, France}

\author{Ludovic Bellon}
\email{ludovic.bellon@ens-lyon.fr}
\affiliation{Univ Lyon, ENS de Lyon, CNRS, Laboratoire de Physique, F-69342 Lyon, France}

\date{\today}

\begin{abstract}
In most systems, thermal diffusion is intrinsically slow with respect to mechanical relaxation. We devise here a generic approach to accelerate the relaxation of the temperature field of a 1D object, in order to beat the mechanical time scales. This approach is applied to a micro-meter sized silicon cantilever, locally heated by a laser beam. A tailored driving protocol for the laser power is derived to quickly reach the thermal stationary state. The model is implemented experimentally yielding a significant acceleration of the thermal relaxation, up to a factor 30. An excellent agreement with the theoretical predictions is reported. This strategy allows a thermal steady state to be reached significantly faster than the natural mechanical relaxation.
\end{abstract}

\maketitle 

\section{Introduction}

In condensed matter, the temperature fields evolve in general on much longer time scales than their mechanical counterparts (such as stress or strain). Indeed, combining the highest thermal diffusivity in solids~\cite{CRCHandbook} ($D\sim\SI{e-4}{m^2/s}$) and the order of magnitude of the speed of sound~\cite{CRCHandbook} ($c\sim[\num{e3}-\num{e4}]\,\si{m/s}$), one defines a typical length scale $l_{th}=D/c\sim[\num{10}-\num{100}]\,\si{nm}$ beyond which temperature is a slow phenomenon. Most systems imply larger length scales, thus a slow temperature evolution. In some devices or experiments, or for proof-of-concept demonstrations, it can however be desirable to accelerate the heat diffusion so as to impose a temperature variation on time scales equivalent, or shorter, than those of the mechanical response of the system. 

One example is atomic force microscopy (AFM)~\cite{binning_atomic_1986}, where a sharp tip attached to a cantilever scans a sample to map its topography and potentially more local properties of its surface. Most commonly, the measure of the interaction force through the deflection of the cantilever is performed with a laser beam~\cite{Meyer-1988}, hence the photothermal response and thermal stability of AFM-sensors have been extensively studied almost since the origin of AFM~\cite{Marti-1992, Allegrini-1992}. Photothermal excitation has been used in vacuum, air, or fluid, aiming at driving the cantilever efficiently while avoiding overheating of the system~\cite{McCarthy-2005, Ramos-2006, lee_thermal_2007, milner_heating_2010, Kiracofe-2011, Bircher-2013}. Let us compare the relaxation time $\tau=L^2/D$ of the temperature field along the cantilever length $L$, to the period of oscillation of the first resonance $T_0\simeq6.4 L^2/(cH)$~\cite{Garcia-2002}, with $H$ the thickness of the cantilever: $T_0/\tau\simeq 6.4\, l_{th}/H=\SI{65}{nm}/H$, where the numerical application has been done for silicon \footnote{Silicon is the most common material for AFM cantilevers}. With $H$ typically in the few $\si{\mu m}$ range, thermal diffusion is thus much slower than the oscillation period of the system, limiting the efficiency of photothermal excitation, or the speed of operation of scanning thermal microscopy~\cite{Chen-2011, gomes_2015, aguilar_sandoval_resonance_2015}. 

Another example where fast temperature variations could be desirable is in stochastic thermodynamics experiments on micro-mechanical systems~\cite{sek10,Rademacher-2022} (optically trapped particles, micro-cantilevers, MEMS - micro-electromechanical systems). When constructing a stochastic Carnot heat engine for instance~\cite{Schmiedl-2007, Toyabe-2010, Esposito-2010, Blickle-2012, Esposito-2012, Holubec-2014, Martinez-2015, Martinez-2016, Plata-2020, DGO22}, one needs to perform adiabatic heating or cooling, that is to say change the system temperature much faster than the time scale $\tau_r$ corresponding to the heat exchanges with the thermostat. $\tau_r$ is equivalent to the mechanical relaxation time, which writes for a cantilever $\tau_r=Q T_0/\pi$, with $Q$ the quality factor of the resonance~\cite{sek10, Dago-2021-PRL, Aurell_2012, Seifert_2012, Jarzynski_2011, Dago-2022-PRL, Dago-2022-JSTAT}. The situation is thus somewhat more favorable than previously for underdamped systems, but requires quality factors larger than a few hundred to start matching the natural thermal diffusion and mechanical relaxation timescales: $\tau_r/\tau \simeq Q \times \SI{20}{nm}/H$. Here again, an acceleration of the temperature dynamics would be welcome to perform efficient adiabatic transformations.

The motivation of this work is to speed up the natural thermal relaxation, and we focus in this article on the case of a cantilever, in relation to the previously mentioned examples. Engineered accelerated dynamics and shortcuts are protocols of the `Shortcut to Adiabaticity' type \footnote{in this expression, adiabatic is to be understood in the sense of a slow transformation}. This class of ideas emerged in the quantum realm~\cite{chen_fast_2010,guery-odelin_shortcuts_2019}. Yet, a number of techniques belonging to this family of accelerating methods has been successfully exported to classical and stochastic dynamics~\cite{guery-odelin_nonequilibrium_2014, martinez_engineered_2016, patra_shortcuts_2017, li_shortcuts_2017,DGO22}. Some of these shortcuts are of inverse engineering type, requiring a careful monitoring of the time dependence of some control parameters, to impose a prescribed evolution of interest~\cite{martinez_engineered_2016}. Such accelerating protocols have also been used fruitfully in systems such as cranes~\cite{2017Gonzalez} and capacitors~\cite{2019PalaiaPhD}. 

In the following, we address the question of the temperature field control, so far untouched. The goal is to drive the thermal relaxation of a micro-cantilever in a time much smaller than its natural relaxation time. First, we model the thermal relaxation and derive theoretically the accelerating protocols allowing the reduction of the relaxation time of the cantilever to reach thermal steady state. Second, the aforementioned protocols are experimentally implemented in a silicon cantilever. The efficacy of the driving is patent when comparing to a direct relaxation process. Moreover, experimental results are in full agreement with the predictions. Finally, we summarize the conclusions of our study and mention future perspectives. 

\section{Theory}\label{sec:Theo}

Our objective is to accelerate the relaxation towards a non-equilibrium steady state of a cantilever with a tunable punctual and well-localized heat source. To this end, we first identify the different eigenmodes (and corresponding timescales) of the heat equation that governs the dynamics of the system. We subsequently design the rate at which the deposit of heat should be carried out so as to cancel the contribution of the low frequency eigenmodes that slow down the relaxation. This original trick enables one to benefit from a fast relaxation with the cancelation of just a few modes. 

\subsection{Temperature rise of a cantilever irradiated by a laser beam}

We gather here the main ingredients useful for a quantitative description of the temperature field of a cantilever irradiated by a laser beam. As the cantilever is placed in vacuum, no heat transfer can occur by convection with its surroundings. Neglecting thermal radiation, the only possible heat transfer mechanism is thus thermal conduction through the cantilever. Considering a cantilever having a length $L$ much larger than its transverse dimensions, the temperature $\theta$ may be assumed homogeneous across the cross section. Therefore, $\theta$ depends only on the longitudinal coordinate $x \in [0,L]$ and time $t$, its dynamics being described by the one-dimensional heat diffusion equation 
\begin{align}\label{eq:heatdiff}
 \rho c_p \frac{\partial \theta(x,t)}{\partial t} &= \lambda \frac{\partial^2 \theta(x,t) }{\partial x^2} + q(x,t),
\end{align}
where $\rho$ is the density, $c_p$ is the heat capacity, $\lambda$ is the thermal conductivity (all of them assumed constant), and $q$ is the heat source/sink density. In vacuum, all cantilever surfaces are assumed to be thermally insulated, except at the location $x_0$ irradiated by the focused laser. We model the heating effect of the focused beam by the source term
\begin{align}\label{eq:sourceterm}
 q(x,t)&=\frac{aP(t)}{S}\delta(x-x_0),
\end{align}
where $a$ is the fraction of light absorbed by the cantilever, $P(t)$ is the incident power of the laser beam which can be manipulated at wish, $S$ is the cantilever cross section area, and $\delta$ Dirac's distribution. 
Furthermore, we assume the following boundary conditions
\begin{subequations} \label{eq:BC}
\begin{align}
 \theta(0,t)&=0,\label{eq:BC0}\\
 \frac{\partial \theta}{\partial x}(L,t)&=0. \label{eq:BCL}
\end{align}
\end{subequations}
These conditions reflect respectively that the cantilever is in contact with the macroscopic chip acting as a thermostat at its clamp $x=0$ (\ref{eq:BC0}), and is isolated on its free end $x=L$ (\ref{eq:BCL}). Note that, for the sake of simplicity, $\theta$ is not the absolute temperature, as we define the origin by the temperature of the chip.

\subsection{Stationary profile \texorpdfstring{$\theta_s$}{}}

The stationary temperature profile $\theta_s(x)$ associated to a given power of the laser $P_f$ is obtained solving Eq.~\eqref{eq:heatdiff} where $\partial \theta/\partial t=0$, with the boundary conditions Eq.~\eqref{eq:BC}. We get
\begin{equation}\label{eq:temp_statio}
\frac{\theta_s (x)}{\theta_s^m} = \left\{
 \begin{array}{ll}
 x/x_0, & 0<x<x_0, \\
 1, & x_0<x<L,
 \end{array}
\right.
\end{equation} 
where $\theta_s^m= a P_f x_0/(S\lambda)$ corresponds to the maximum temperature elevation. Between the chip and the laser spot ($x=x_0$), the stationary temperature increases linearly with the position $x$, while beyond the laser spot, it remains constant and equals $\theta_s^m$.

From now on, we focus the analysis on the relaxation between two stationary states when driving the power $P(t)$ from a constant initial value $P(t < 0)=P_0$ to its constant final value $P(t > t_f)=P_f$. Note that even if $P(t)$ is constant after the final time $t_f$, the temperature profile continues to evolve towards its final asymptotic stationary shape. Since the heat diffusion equation (Eq.~\ref{eq:heatdiff}) is linear, we make the simplifying assumption that $P_0=0$: without loss of generality, $P(t)$ or $P_f$ represent the laser power excess with respect to $P_0$, and $\theta(x,t)$ or $\theta_s(x)$ stand for the temperature excess with respect to the initial stationary temperature profile.

\subsection{Transient solution through expansion}

The general solution of Eq.~\eqref{eq:heatdiff} can be expressed carrying out an expansion in eigenfunctions $\psi_n(x)$ \cite{Xiong_Hilbert_2017,Krivtsov_Ballistic_2019}, each satisfying the associated homogeneous problem $\partial_{\Tilde{x}}^2 \psi_n(\Tilde{x})=-k_n^2\psi_n(\Tilde{x})$ with $\psi_n(0)=0$ and $\psi_n'(L)=0$. Solving the homogeneous problem leads to
\begin{align}
 \psi_n(\Tilde{x})=\sqrt{2}\sin \left(k_n \Tilde{x} \right), \quad \Tilde{x}=x/L
\end{align}
with the eigenvalues $k_n=(n-1/2)\pi$, $n \in \mathbb{N}^*$. The temperature profile can be expressed as 
\begin{equation}\label{eq:temp_xt}
 \theta(x,t)=\theta_s(x) + \theta_s^m \sum_{n=1}^\infty C_n(\Tilde{t}) \psi_n( \Tilde{x}),
\end{equation}
where the components $C_n(\Tilde{t})$ depend solely on $\Tilde{t}=t/\tau$, with $\tau=\rho c_p L^2/\lambda$, the characteristic diffusion time of the problem. These components are obtained projecting Eq.~\eqref{eq:heatdiff} on the eigenmodes basis $\psi_n(x)$; they read as
\begin{align}\label{eq:Cn}
 C_n(\Tilde{t})&=\frac{\psi_n(\Tilde{x}_0)}{\Tilde{x}_0}\left[\int_0^{\Tilde{t}} \big(F(s\tau)-1\big) e^{ k_n^2s}ds-\frac{1}{k_n^2}\right]e^{-k_n^2\Tilde{t}},
\end{align}
with $\Tilde{x}_0=x_0/L$, and $F$ the driving function defined as 
\begin{equation}
 F(t)=\frac{P(t)}{P_f}.
\end{equation}
The temperature solution Eq.~\eqref{eq:temp_xt} is completely general for arbitrary $F(t)$, even though here $F(t<0)=0$ and $F(t>t_f)=1$ by construction. For $t>t_f$, the bracket term in Eq.~(\ref{eq:Cn}) is constant, so each component in the expansion decays exponentially with its own decaying rate. Specifically, the $n$-th component has a rate equal to $k_n^2$, which increases with the mode number $n$ as $(2n-1)^2$: the second mode decays $k_2^2/k_1^2=9$ times faster than the fundamental mode, the third mode decays $k_3^2/k_1^2=25$ times faster than the fundamental one, etc. The higher the mode $n$, the less time it takes for that component of the temperature to relax.
 
\subsection{Accelerating the dynamics}

\begin{figure}[tb]
 \centering
 \includegraphics{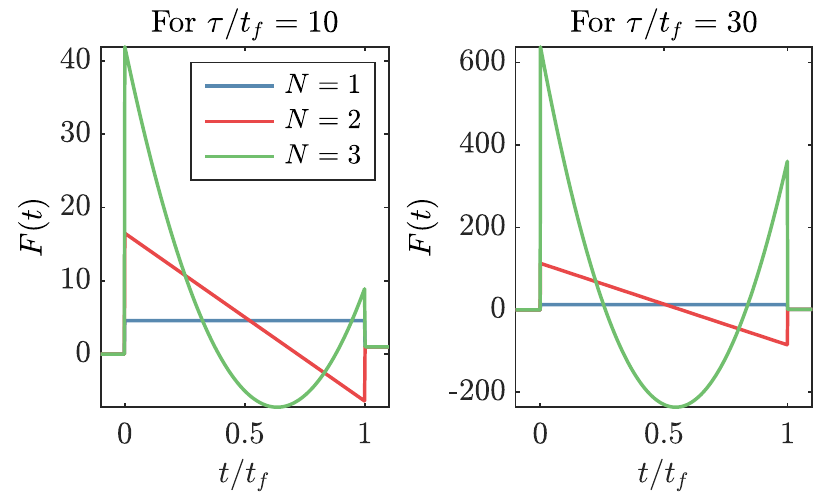}
\caption{\label{fig:Ffunction}Driving function $F(t)$ to apply in order to cancel successively the first thermal modes for an acceleration (towards the stationary profile) such that $t_f=\tau/10$ (left) and $t_f=\tau/30$ (right). The coefficients $\gamma_n$ defining $F(t)$ are found solving Eq.~\eqref{eq:systemGamCoef}; they are listed in table~\ref{table:gamma} of Appendix \ref{appendix:gamma}.}
\end{figure}

We explain hereafter how the driving function can be designed to remove the contribution of the slow relaxation modes,
thereby speeding up the relaxation dynamics in a given arbitrary small (chosen) time lapse $t_f$. Our strategy is somewhat similar to that of Ref. \onlinecite{Gal-2020} on a different thermal acceleration problem. For this purpose, we use an ansatz for $F(t)$ that involves as many parameters as modes to be canceled. Let us assume that we want to cancel the $N$ first modes. We can take the following convenient polynomial ansatz
\begin{equation}\label{eq:protocfct}
F(t) = \sum_{m=1}^N\gamma_m \left(t/t_f\right)^{m-1}, \quad 0<t<t_f.
\end{equation}
Canceling the $N$ first modes corresponds to impose $C_n(\Tilde{t}_f)=0$ for $n=1$ to $N$, where $\Tilde{t}_f=t_f/\tau$. This implies for $n\in[1,N]$
\begin{align}\label{eq:systemGamCoef}
 \sum_{m=1}^N \gamma_{m}\int_0^{\Tilde{t}_f} \left( \frac{s}{\Tilde{t}_f}\right)^{m-1} e^{ k_n^2s} ds = \frac{e^{k_n^2\Tilde{t}_f}}{k_n^2} 
\end{align}
which is a system of $N$ linear equations for $N$ unknown variables $\gamma_n$, analytically solvable. When $t_f$ is reached, we change the laser power abruptly to $P_f$ (i.e. $F(t>t_f)=1$) and the relaxation will occur only through the modes higher than the $N$-th one (since we already force the cancelation of the first $N$ modes). Examining Eq.~\eqref{eq:systemGamCoef}, the coefficients $\gamma_n$ depend only on the normalized protocol duration $\Tilde{t}_f$ and the number of canceled modes $N$; the protocol to apply thus does not depend on the laser beam position $\Tilde{x}_0$. For the sake of concreteness, we illustrate our protocols with $\Tilde{x}_0=0.95$ (we  exemplify the independence of the protocol on $\Tilde{x}_0$ by working out the solution for $\Tilde{x}_0=0.5$ in Appendix \ref{appendix:x0}). If one wants only to cancel the relaxation of the fundamental mode ($N=1)$, the function $F(t)$ to apply during the protocol is constant and equals $1/(1-e^{-k_1^2\Tilde{t}_f})$; for $\Tilde{t}_f=0.1$, we have $F(t)=4.57$. In Fig.~\ref{fig:Ffunction}, we show the functions $F(t)$ that allow the cancelation of up to the third mode for a protocol duration $\Tilde{t}_f=1/10$ and $\Tilde{t}_f=1/30$. Further details on the computation and properties of $F(t)$ are given in Appendix \ref{appendix:gamma}. The needed range of $F(t)$ (thus the laser power) increases with both the number of canceled modes $N$ and the acceleration factor $\tau/t_f$. Note that for $N>1$, $F(t)$ requires both positive and negative values. In practice, the cantilever can only absorb the power from the laser beam. However, $P(t)$ corresponds to the power excess with respect to the initial value $P_0$, so that the protocol will be valid as long as $P_0$ is larger than $-\min[P(t)]$. A priori, this constraint implies a  limitation for our protocols. However, we prove below that our method succeeds in accelerating thermal relaxation substantially in a wide variety of experimentally realizable situations.

\section{Experimental results}
\subsection{Experimental setup}

For the experiments presented below, a silicon cantilever placed in vacuum is heated with a laser and the evolution of its temperature at various positions $x$ along its length $L$ is measured.

At a thickness of a few micrometers, silicon is semi-transparent for visible light, the light experiences multiple reflections within the thickness of the cantilever, which acts as a lossy Fabry-Perot~\cite{pottier_thermo-optical_2021}. As a result, the fraction of light reflected $R$ depends on the thickness and the silicon refractive index, both varying  with temperature $\theta$. By measuring the local change in reflectivity, one can thus infer the local temperature variation with respect to the reference temperature field. For small variations, we have $ \Delta R(x)/R(x)= \beta \theta(x)$, where the sensitivity $\beta$ depends on the local properties (refractive index and thickness). The calibration procedure for $\beta$ is detailed in Appendix \ref{appendix:beta}.

\begin{figure}[tb]
 \centering
 \includegraphics[scale=0.31]{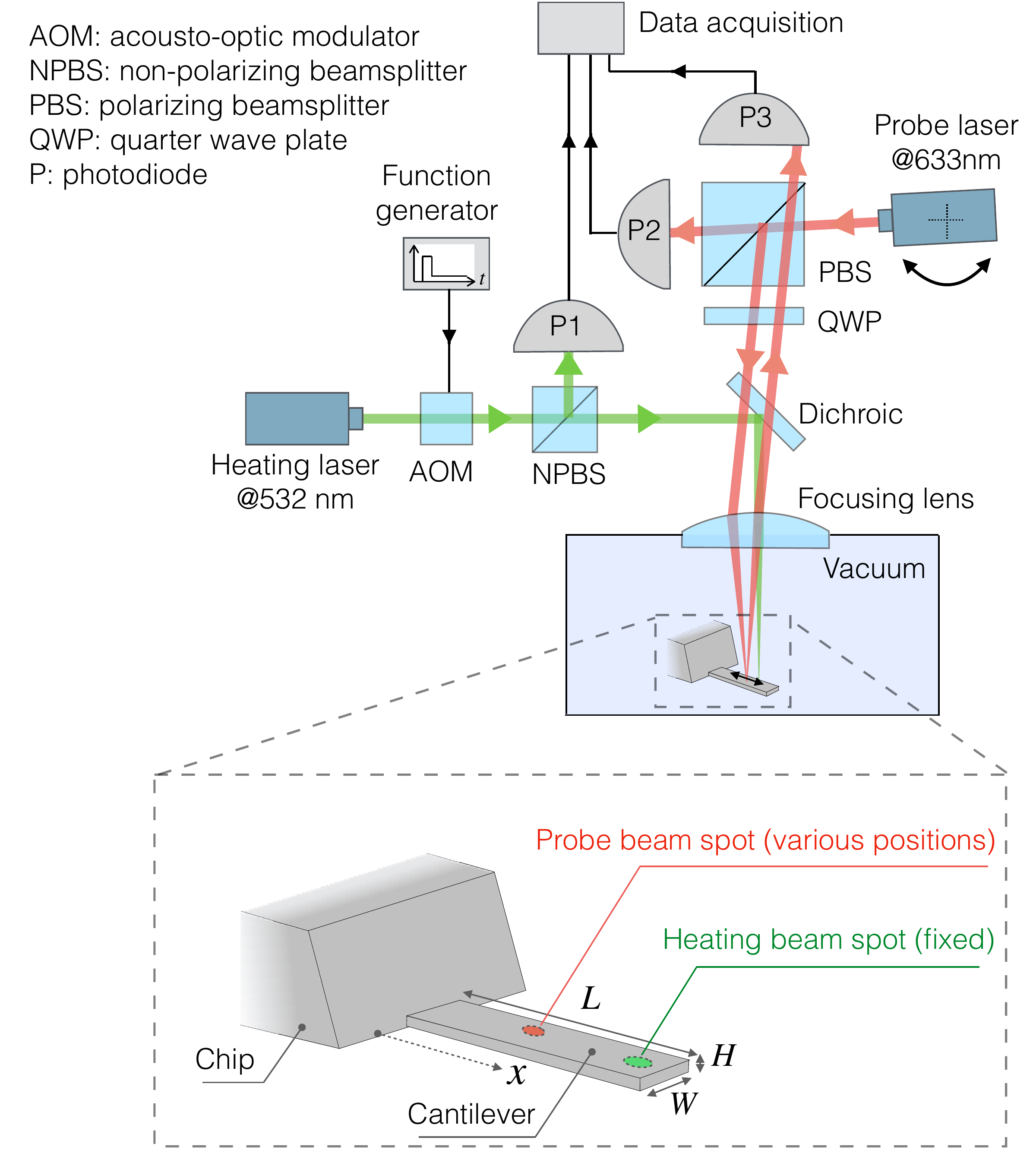} 
\caption{\label{fig:SetupExperiment} Sketch of the experimental setup. The heating laser beam (at $\SI{532}{nm}$) is fixed and focused at the extremity of a cantilever placed in vacuum. The power of the heating beam is modulated with an acousto-optic modulator. The probe laser beam (at \SI{633}{nm}) is used to measure the local temperature change by analyzing the cantilever reflectivity variations. By tilting the probe laser, the probe beam spot scans the cantilever length allowing the measurement of the temperature variation at different positions $x$.}
\end{figure}

The experimental setup comprises two laser beams (Fig.~\ref{fig:SetupExperiment}): (i) a heating laser beam (at 532~nm) of variable power $P(t)$ focused at a fixed position close to the cantilever free end ($x_0=0.95L$) that serves to heat the cantilever, and (ii) a probe laser beam (at 633~nm) of constant power ($\SI{350}{\mu W}$) that allows us to measure the local temperature rise taking benefit of the thermo-optical effect described above. The radius of the heating and probe beam at the cantilever surface are respectively $\SI{14}{\mu m}$ and $\SI{7}{\mu m}$. The power of the heating beam is modulated with an acousto-optic modulator, with a rise time of $\SI{100}{ns}$. Three photodiodes $P1$, $P2$, and $P3$ measure the incident heating beam power $P(t)$ and the reflectivity $R$ of the probe beam respectively, as shown in Fig.~\ref{fig:SetupExperiment}. By tilting the probe laser, one scans the cantilever length facilitating temperature measurements at various locations $x$. 

During the experiment, the cantilever is placed in vacuum at \SI{2e-2}{mBar}. At this pressure level, the contribution of convective heat transfer is negligible compared to thermal conduction~\cite{lee_thermal_2007}. All the measurements are performed using a single raw silicon cantilever OCTO-1000D from Micromotive, with the following size: length $L=\SI{1}{mm}$, width $W=\SI{90}{\mu m}$ and thickness $H=\SI{5}{\mu m}$. 

\begin{figure*}[tb]
 \centering
 \includegraphics{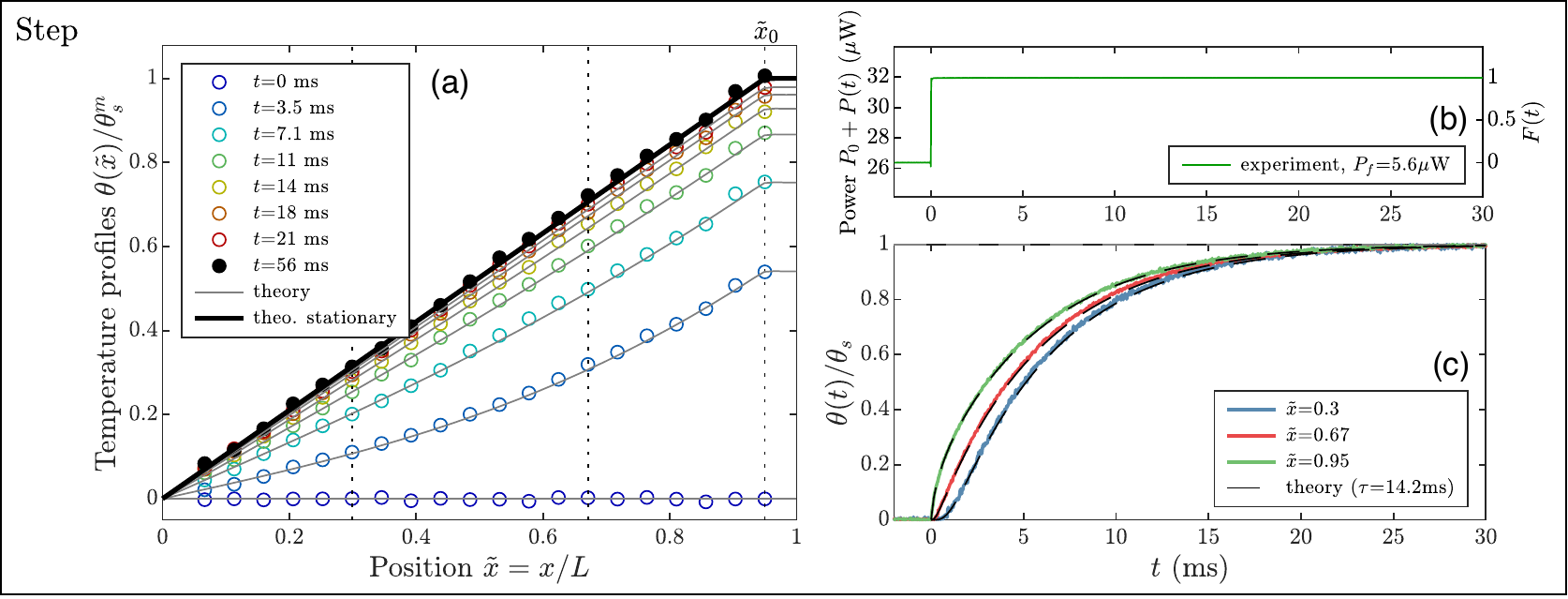} 
\caption{\label{fig:step}Transient temperature profiles (a) of a cantilever heated by a laser whose incident power $P(t)$ is the step function (b). The heating laser is located near the cantilever free end ($\Tilde{x}_0=0.95$). The $\SI{5.6}{\mu W}$ increase in power induces a maximum temperature elevation of $ \theta_s^m=\SI{38}{mK}$. As time goes on, the temperature profile converges towards the expected linear stationary profile given by Eq.~\eqref{eq:temp_statio}. Graph (c) shows the normalized evolution of temperature at the positions $\Tilde{x}=0.3$, 0.67 and 0.95,
corresponding to the vertical dashed lines in panel a). All experimental data are perfectly described by the model of Eq.~\eqref{eq:temp_xt}, using the characteristic time $\tau=\SI{14.2}{ms}$.}
\end{figure*}

To record the spatial dependence of the temperature profile, we choose 20 evenly spaced positions $x$ of the probe beam along the cantilever. At each position, we measure the time evolution of the temperature $\theta(t)$ induced by the applied incident power $P(t)$ of the heating beam. To increase the signal-to-noise ratio, the displayed temperature variations are obtained averaging over 700 heating procedures at each $x$. The spatial resolution is approximately the size of the laser spot, $\SI{7}{\mu m}$, hence smaller that $\SI{1}{\%}$ of the cantilever length.

\subsection{Relaxation for a step function (\texorpdfstring{$F(t)=1$}{})} 

Before testing the performance of the accelerating protocols worked out in section~\ref{sec:Theo}, we first present the results obtained imposing a jump in power, $P(t)=P_f$ for $t>0$. The analysis of the temperature relaxation allows measuring the characteristic diffusion time $\tau$ needed to determine the function $F(t)$, to later successfully accelerate the dynamic given a protocol duration $t_f$. Moreover, this is the simplest reference that can be thought of and it defines the timescale that we would like to beat.

In Fig.~\ref{fig:step} and in the movies available as ancillary files~\cite{suppmat-movies}, we display the cantilever temperature rise measured imposing a sudden increase in the power of $P_f=\SI{5.6}{\mu W}$. The heating laser spot is focused close to the free end of the cantilever at $x_0=0.95L$. The convergence of the temperature profile towards the expected linear stationary profile given by Eq.~\eqref{eq:temp_statio} confirms that for the temperature variations explored, the silicon conductivity $\lambda$ can be assumed constant. The maximum temperature elevation is $\theta_s^m=\SI{38}{mK}$, this value is consistent with the theoretical one $\theta_s^m=a P_f x_0/(WH\lambda)$ with $\lambda=\SI{156}{W.m^{-1}.K^{-1}}$~\cite{masolin_thermo-mechanical_2013} and $a=0.5$, an absorption coefficient also consistent with the $H=\SI{5}{\mu m}$ thickness of the cantilever~\cite{pottier_thermo-optical_2021}.

\pagebreak

In the case of a step function ($F(t)=1$), the temperature is predicted by Eq.~\eqref{eq:temp_xt} where the components $C_n(\Tilde{t})$ are reduced to \begin{align}\label{eq:c_step}
 C_n(\Tilde{t})&=- \frac{ \psi_n(\Tilde{x}_0) }{ \Tilde{x}_0 } \frac{1}{k_n^2} e^{-k_n^2\Tilde{t}}.
\end{align}
Using Eq.~\eqref{eq:c_step}, all experimental data are properly described by theory for a characteristic time $\tau$=\SI{14.2}{ms}. From the thermal diffusivity of bulk silicon given in the literature~\cite{masolin_thermo-mechanical_2013}, $D=\lambda/\rho c_p=\SI{86}{mm^2/s}$, we expect $\tau=L^2/D=\SI{11.6}{ms}$. The 18\% difference could be explained by the reduced diffusivity of silicon due to the phonon confinement effect. Indeed, for a $\SI{5}{\mu m}$ thick film, the silicon conductivity is expected to be approximately 15\% smaller than the value for bulk silicon~\cite{wang_computational_2014}. 

\subsection{Relaxation for an acceleration factor \texorpdfstring{$\tau/t_f=10$}{of 10} }

In Fig.~\ref{fig:AllProtocol}, and in the movies available as ancillary files~\cite{suppmat-movies}, we display the evolution of the measured cantilever temperature when applying the driving function $F(t)$ given by Eq.~\eqref{eq:protocfct} for a protocol duration $t_f=\SI{1.4}{ms}$, corresponding to $\Tilde{t}_f=0.1$. We test our protocol varying the number of canceled modes $N$ from 1 up to 4. Note that the imposed functions $F(t)$ correspond to the ones already presented in Fig.~\ref{fig:Ffunction} (left panel). In order to be able to impose the negative variations in $F(t)$ (for $N>1$), we choose a reference incident power of $P_0=\SI{210}{\mu W}$. As the number of canceled modes $N$ increases, the temperature at the end of the protocol ($t=t_f$) clearly converges towards the stationary profile $\theta_s(x)$ (Fig.~\ref{fig:Conv_comp}). Contrary to the step function (Fig.~\ref{fig:step}), the temperature takes intermediate values higher than the stationary target $\theta_s(x)$. The maximum transient temperature is obtained for $N=4$ and corresponds to $\theta=4.7 \times \theta_s^m$ at $x=0.95L$. For each protocol, we compare the experiment to theory computing the temperature from Eq.~\eqref{eq:temp_xt} limiting the infinite sum to 60 terms; all experimental data remarkably coincide with theory.

\setcounter{figure}{4}
\begin{figure}[!bht]
 \centering
 \includegraphics{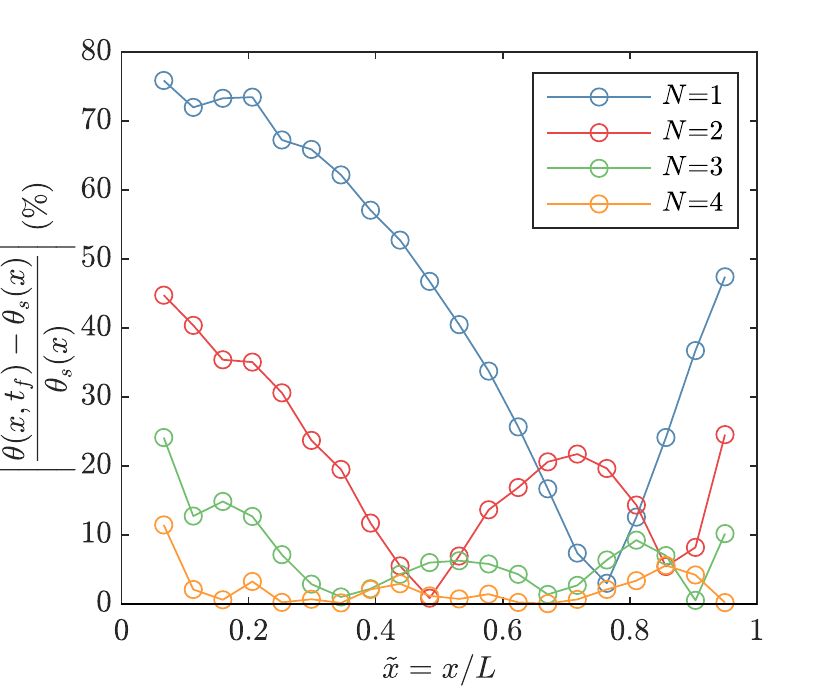}
\caption{\label{fig:Conv_comp}Relative temperature difference from the targeted stationary profile $\theta_s(\tilde{x})$ at the end of the protocol $t=t_f=0.1\tau$. As the number of canceled modes $N$ increases, the temperature profile converges towards the stationary profile. For $N=4$, the relative temperature difference remains below 10\% along the cantilever length.}
\end{figure}

\setcounter{figure}{3}
\begin{figure*}[htb]
 \centering
 \includegraphics{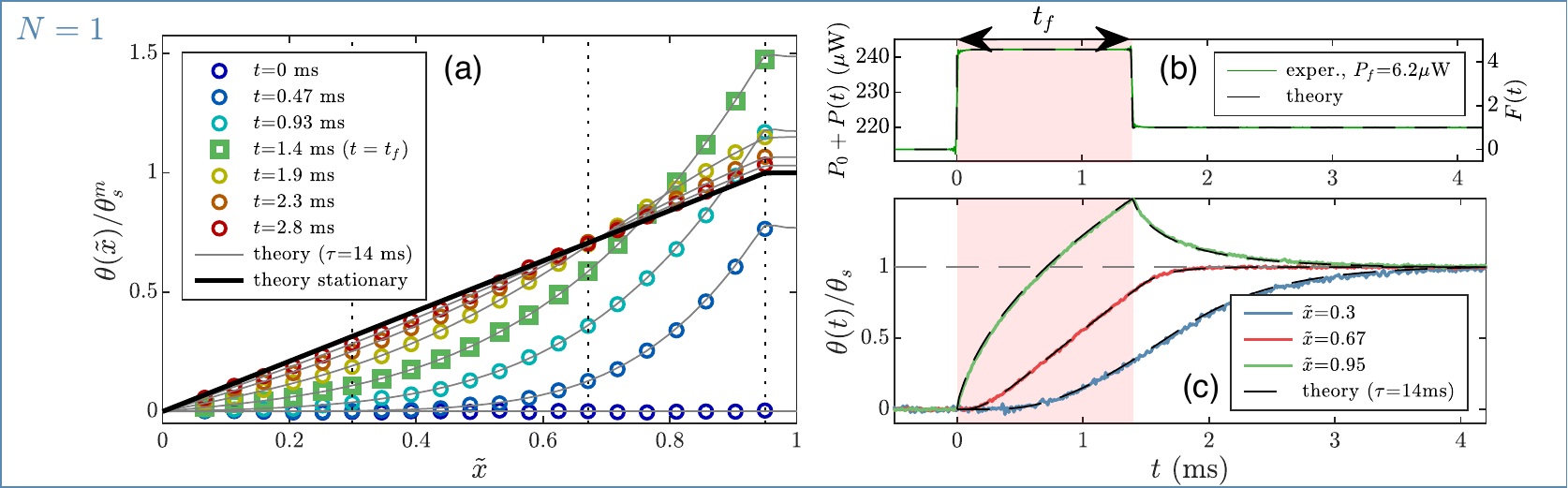} \\
 \vspace{1mm}
 \includegraphics{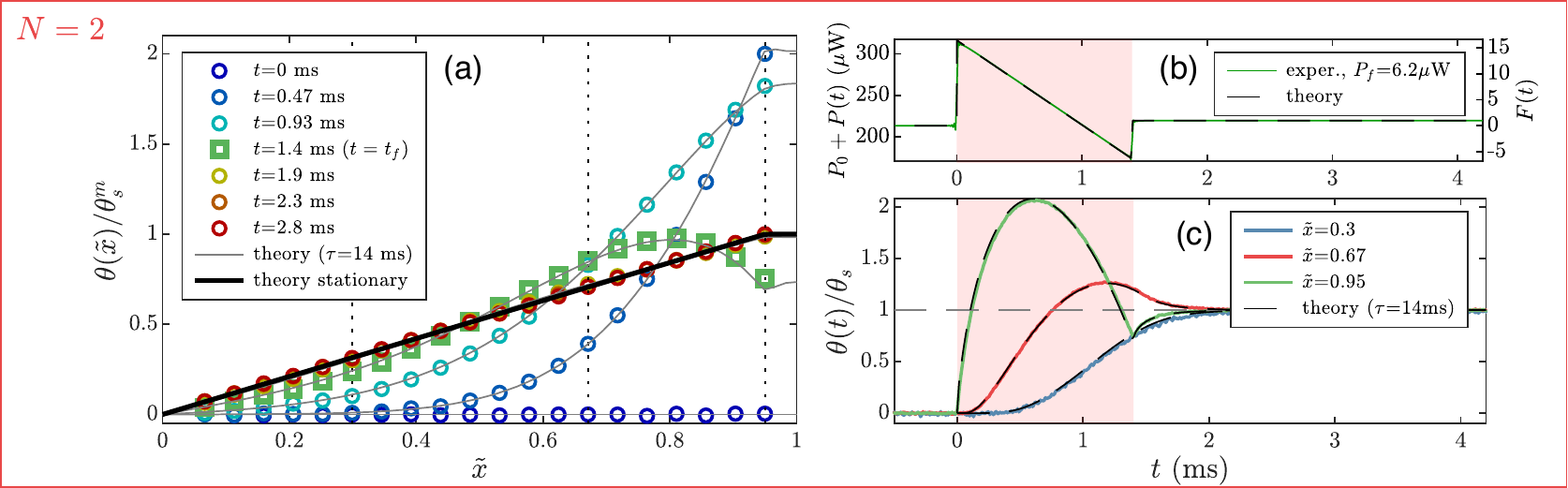} \\
 \vspace{1mm}
 \includegraphics{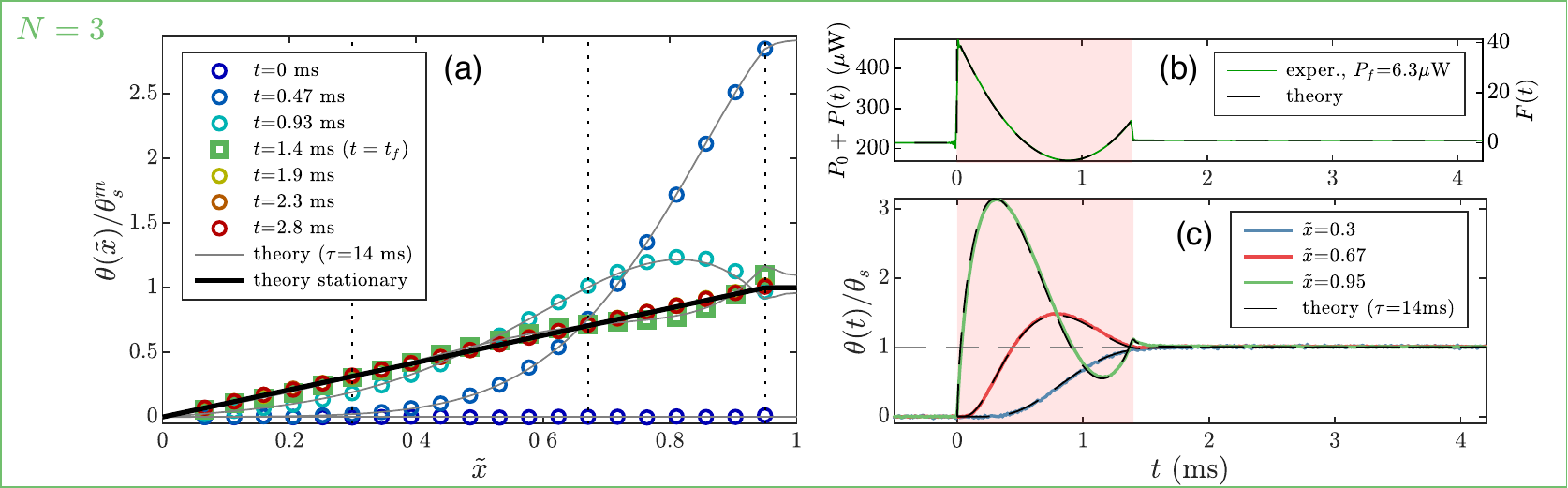} \\
 \vspace{1mm}
 \includegraphics{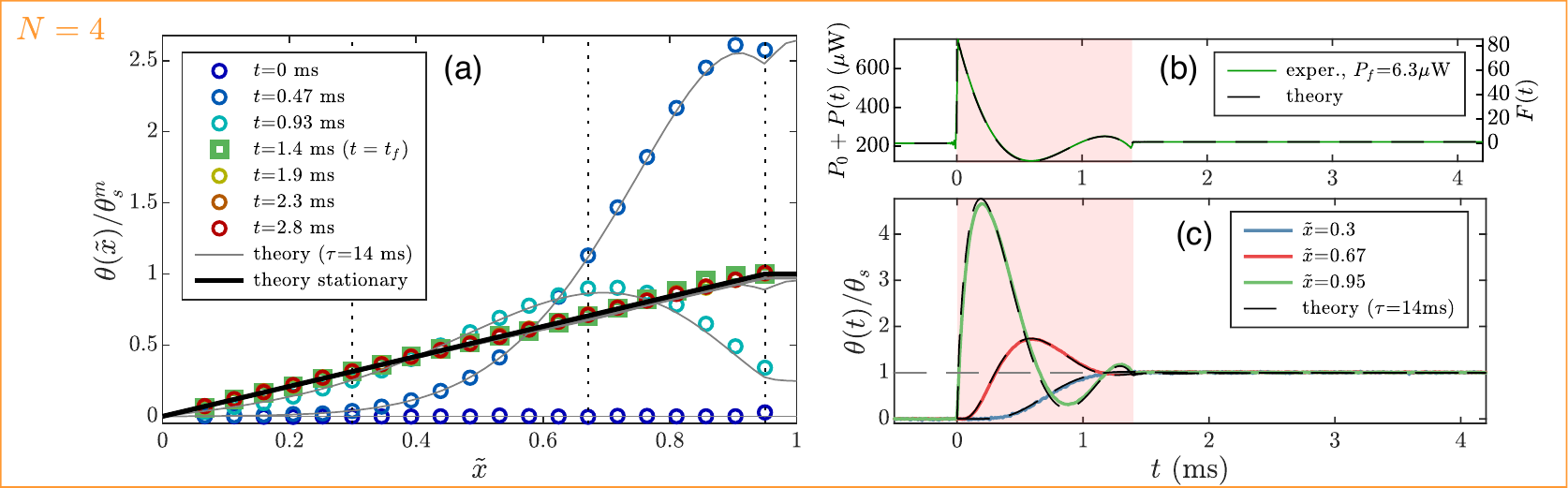} 
\caption{\label{fig:AllProtocol}Transient temperature profiles measured when applying the protocol function $F(t)$ given by Eq.~\eqref{eq:protocfct} with a protocol duration $t_f=0.1\tau$ designed to cancel the $N$ first thermal modes. For each tested protocol ($N=1$ up to $N=4$) we display (a) the temperature profiles, (b) the imposed power $P(t)$, and (c) the temperature variations with time at the positions $\Tilde{x}=0.3$, 0.67 and 0.95. All experimental data coincide with the expected theoretical temperature variations.}
\end{figure*}
\setcounter{figure}{5}

\pagebreak

\begin{figure}[ht]
 \centering
 \includegraphics{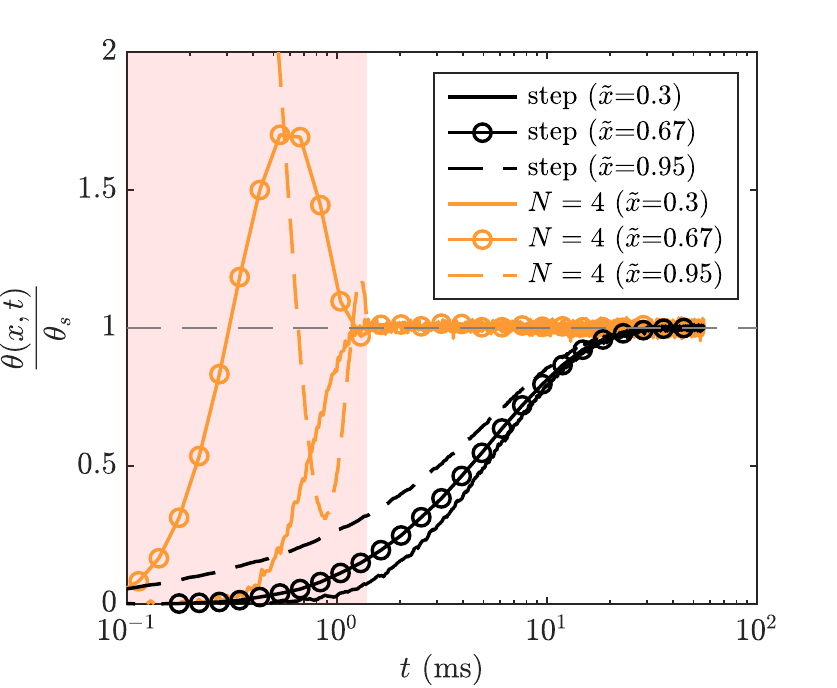}
\caption{\label{fig:Conv_log}Comparison of the temperature convergence towards the stationary profile $\theta_s$ when heating the cantilever with a step function power (black) or with the accelerating protocol function of Eq.~\eqref{eq:protocfct} for $N=4$ (orange). The red shaded area corresponds to the protocol duration, $t_f=0.1\tau=\SI{1.4}{ms}$.}
\end{figure}
    
To highlight the success of our tailored acceleration, we compare in Fig.~\ref{fig:Conv_log} the difference of convergence towards the stationary profile $\theta_s$ between the step function power (displayed in Fig.~\ref{fig:step}) and the accelerating protocol for $N=4$ (displayed in the last panel of  Fig.~\ref{fig:AllProtocol}). Using the step function, the duration needed to converge within 2\% (at the positions $\Tilde{x}=0.3,0.67$ and 0.95) is $\SI{25}{ms}$ (1.8$\tau$), while in the accelerating protocol this milestone is reached at the end of the program $t_f=\SI{1.4}{ms}$ (0.1$\tau$), beating remarkably the natural relaxation time. Actually, having gotten rid of the first four modes, the fifth and higher orders do remain and are responsible for a slight mismatch with respect to the target profile. They decay at least $9^2=81$ times faster than the power step-forcing relaxation, with time scales smaller or equal to $\tau/(4.5\pi)^2 \simeq \SI{0.07}{ms}$, much smaller than $t_f$ itself. Hence, since the characteristic evolution timescale (after $t_f$) is $k^{-2}_{N+1} \tau$, high values of $N$ are needed  for extremely low values of $t_f/\tau$ only. 

\subsection{Relaxation for an acceleration factor \texorpdfstring{$\tau/t_f=30$}{of 30} }

\begin{figure}[!tb]
 \centering
 \includegraphics{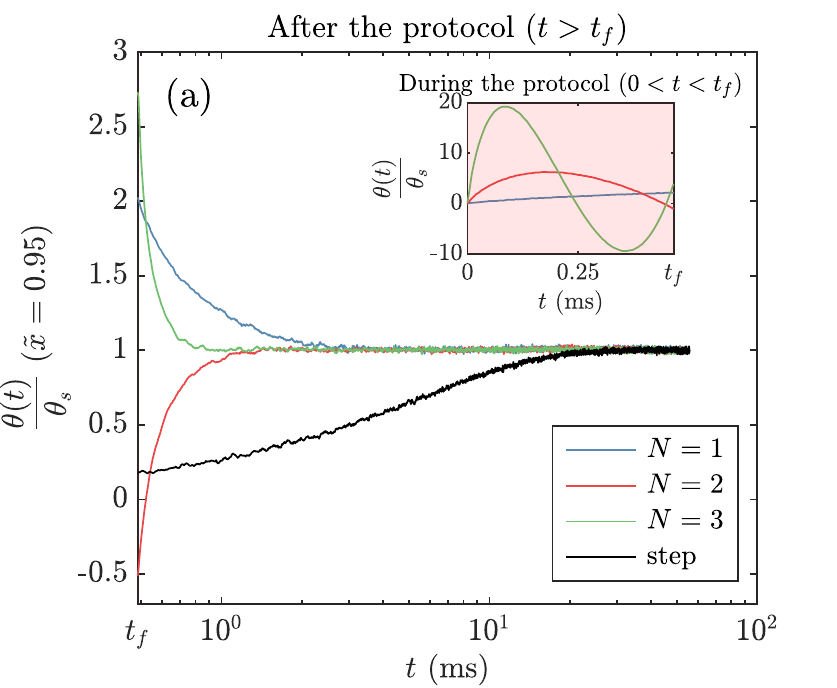}
 \includegraphics{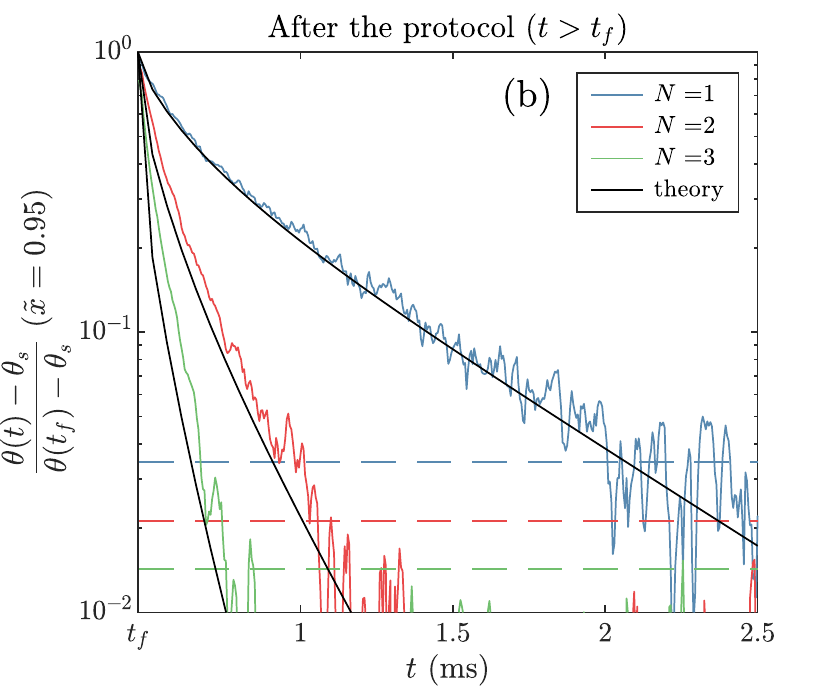}
\caption{\label{fig:acce30}(a) Temperature relaxation towards the stationary state (measured close to the cantilever free end $\Tilde{x}=0.95$) varying $N$ for a protocol duration $t_f=\tau/30=\SI{0.47}{ms}$. Inset: temperature variations during the protocol. (b) Comparison of the temperature decay. The horizontal dashed lines correspond to a relative difference $(\theta-\theta_s)/\theta_s$ of 4\%.}
\end{figure}

In Fig.~\ref{fig:acce30}(a), we compare the temperature convergence towards the stationary state varying $N$ for a shorter protocol duration of $\SI{0.47}{ms}$ corresponding to $\tau/t_f=30$. The driving functions applied, $F(t)$, correspond to the ones displayed in Fig.~\ref{fig:Ffunction} (right panel). To be able to impose the large variations in $F(t)$, a reduced target power change $P_f=\SI{0.6}{\mu W}$, implying a maximum temperature elevation of $\theta_s^m=\SI{5}{mK}$, is considered. To maintain the signal-to-noise ratio, each curve displayed is now obtained averaging over $3000$ measurements. As observed previously for the slower acceleration of $t_f=\tau/10$, the temperature variations during the protocol duration (see inset) increase with the number of canceled modes $N$. Note that the transient temperature variations are now much larger and can take negative values.

In that case, the temperature at the end of the protocol is notably different from the stationary value $\theta_s$. The relative difference $(\theta(t_f)-\theta_s)/\theta_s$ (measured at $\Tilde{x}=0.95$) is respectively 100\%, -150\% and 170\% for $N$ increasing from 1 to 3. It may seem counter-intuitive that the difference from the target value obtained at the end of the protocol increases with the number of canceled modes. We recall that our protocol allows converging towards the stationary profile at the end of the protocol only in the limit $N\rightarrow \infty $. For a finite value $N$ the use of the protocol guarantees that starting from $t_f$ only the modes higher than $N$ will relax. Because the rate of relaxation (equal to $k_n^2\tau$) increases with the mode number $n$, increasing $N$ should allow a faster convergence towards the stationary profile. It cannot be guaranteed that the coefficients of the modes higher than the $N$-th one take small values at $t=t_f$, but its relaxation is ensured to be exponential, faster and faster for increasing $n$. Numerical resolution of the heat equation with our protocols are provided in Appendix \ref{appendix:transient}, quantifying the limitation caused by these modes surviving after $t_f$. In Fig.~\ref{fig:acce30}(b), we verify that the temperature starting from $t_f$ relaxes faster as $N$ increases, as predicted. For $N=3$ for example, the relaxation times are at most $\tau/(3.5\pi^2) \simeq \SI{0.12}{ms}$, smaller than $t_f$ itself, and in agreement with the observation (green curve).

\section{Discussion and Conclusions} 

In this work, we tailor in time the power of a localized irradiating laser, in order to speed up the thermal relaxation of a micro-cantilever in a duration much smaller than its natural relaxation time. The protocol duration can be reduced, in principle, as much as desired. In practice however, a smaller duration unavoidably translates into a larger dynamics of the heating/cooling power (minimum and maximum power with respect to final value, faster time control of the imposed drive), and of the transient values of the temperature field. Experimentally, acceleration factors up to a factor 30 are demonstrated.

The theoretical approach relies on canceling the first $N$ eigenmodes of the thermal diffusion equation in a finite time. Of course, the system is getting closer to the exact steady states as more $N$ are considered. Nevertheless, we prove that for an experimentally accessible range of parameters, relatively small values of $N$ may suffice,  $N=4$ providing a \textit{final} state already very close to the stationary one. We emphasize that such a strategy to speed up a given dynamics described by partial differential equations is generic, and can be applied to a wide variety of mesoscopic systems \cite{2019PalaiaPhD}. It complements the methods that have been recently developed in stochastic thermodynamics to speed up a relaxation process~\cite{DGO22}. Our approach requires that local equilibrium holds at all times, which allows us to define the temperature field and ensure the validity of Eq.~\eqref{eq:heatdiff}. The specific case worked out here also requires that a 1D description of the system is valid, thus that the accelerated dynamics is still slow with respect to the transverse directions of the cantilever. With a width to length ratio of $W/L=0.09$ in our experiments, the transverse relaxation time is $(W/L)^2\tau=8.10^{-3}\tau$, smaller than the smallest $t_f$ we probe, but it could be an issue for more ambitious accelerations !
 
Some driving protocols require a heat sink (negative power), which is technically impossible with laser absorption. From the experimental point of view, this represents a technical problem that is bypassed by choosing a heated initial state. This trick allows the redefinition of the driving function with an offset that gives access to effective negative powers. With our notation, $P_0 + P(t)$ needs to remain positive. Our solution is proven to work in all the considered processes.

Let us point out that our approach is quite different from classic feedback control methods. Those would typically rely on measuring the temperature in one position, say $x_0$, and adjusting the heating power to reach the setpoint, $\theta_s^m$. A perfect feedback loop would therefore achieve $\theta(x_0,t>0)=\theta_s^m$, acting as an effective boundary condition (a temperature step at the specific point $x_0$) for the heat equation \eqref{eq:heatdiff}. The typical time scale to reach the stationary state would then be $x_0^2/(\pi^2 D)$ (see Appendix \ref{appendix:tempstep}), close to one fourth of the power step slowest decay time $4 L^2/(\pi^2 D)$: only a 4 times acceleration of the global dynamics would consequently be achieved. Note that reaching the stationary value in one point is very different from reaching the final profile at every position. As a further illustration, we superimpose in the movies available as ancillary files~\cite{suppmat-movies} the time evolution of the temperature field in response to a power step, to a temperature step (perfect feedback loop), and to our protocols. The acceleration provided by our strategy is clear, at the expense of noticeable local temperature oscillations and overshoots.

\begin{figure}[tbh]
 \centering
 \includegraphics{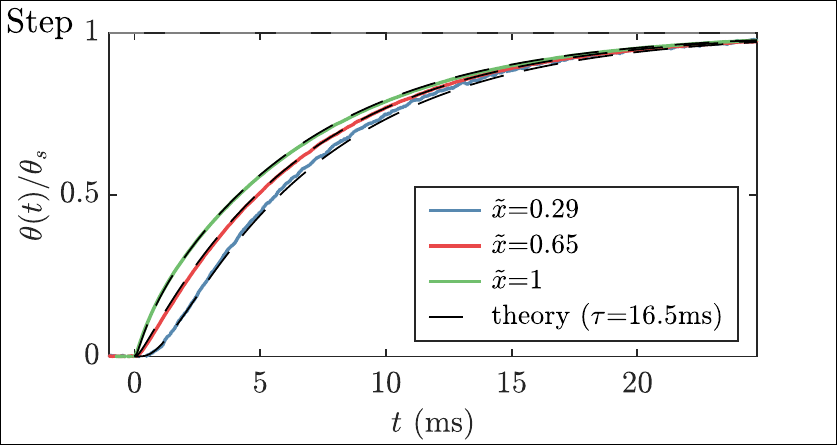}
 \includegraphics{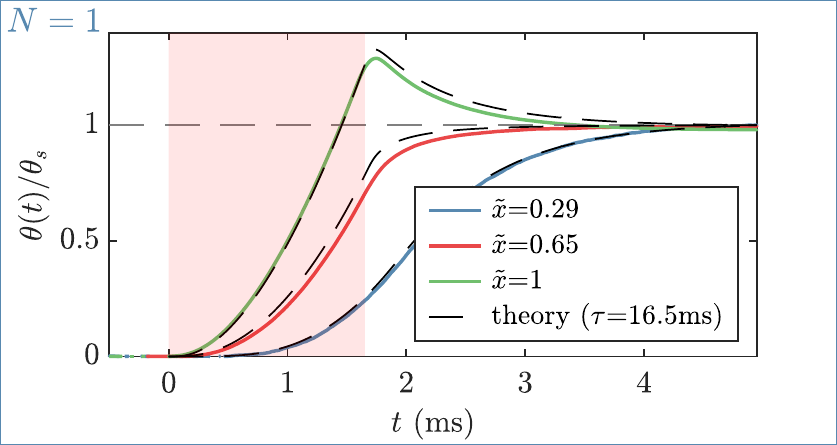}
 \includegraphics{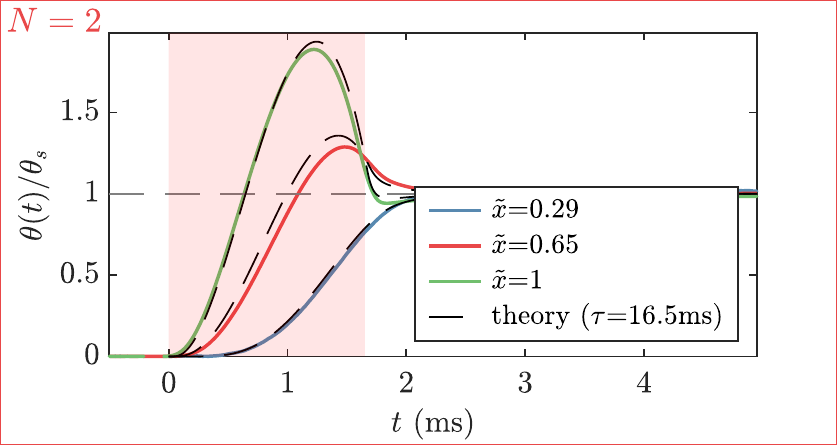}
 \includegraphics{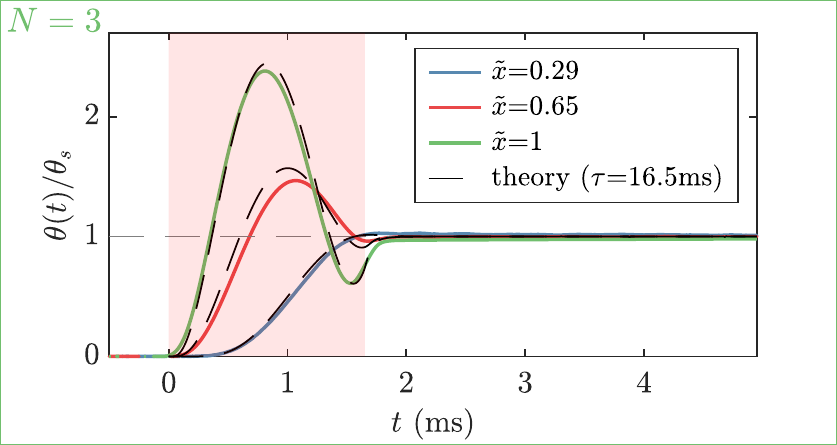}
\caption{\label{fig:testhighertemp}Transient temperature profiles measured when applying a step function and the accelerating protocol ($N=1$ up to $N=3$) for a maximum temperature elevation $\theta_s^m=\SI{1}{K}$. The small deviation from theory is attributed to the thermal conductivity dependence of silicon with temperature: as the initial profile corresponds to a maximum temperature rise of $\SI{50}{K}$, $\lambda$ is up to $\SI{16}{\%}$ smaller near the cantilever's free end.}
\end{figure}

The order of magnitude of the temperature jumps that we experimentally perform is modest: $\SI{38}{mK}$ for a 10 times acceleration, and $\SI{5}{mK}$ for a 30 times acceleration. This choice hinges upon a practical consideration: linearity and dynamic range. The thermal conductivity of silicon depends on temperature in a noticeable way (roughly as $\lambda\propto 1/T$~\cite{masolin_thermo-mechanical_2013}), so that we would like to avoid large temperature excursions to match the linearity hypotheses of the model. Moreover, we need an initial power $P_0$ to allow negative values for $P(t)$, hence the initial temperature profile is actually not flat. For the set of data presented in Figs.~\ref{fig:step} to \ref{fig:Conv_log}, the maximum temperature due to $P_0=\SI{210}{\mu W}$ and the $\SI{350}{\mu W}$ of the red laser probe beam is $\SI{4}{K}$, corresponding to a variation of $\lambda$ below $\num{1.3}{\%}$. Larger $P_0$, allowing larger $P_f$, would imply non uniform values for $\lambda$, again deviating from the model. Our goal in this article is to demonstrate the excellent agreement between theory and experiment, so we restrict ourselves to small temperature jumps. We also test larger steps corresponding to $\theta_s^m=\SI{1}{K}$, and as illustrated in Fig.~\ref{fig:testhighertemp}, the protocols work fine with only tiny adjustments to the parameter $\tau$ to compensate for emerging non-linearities. In this case, the maximum temperature due to $P_0=\SI{400}{\mu W}$ and the $\SI{600}{\mu W}$ of the red laser probe beam is $\SI{50}{K}$, corresponding to a non-uniformity of $\lambda$ of $\SI{16}{\%}$. 

Larger jumps would require the simple form of the dynamic evolution that facilitates the theoretical derivation of the driving protocol to be reconsidered. The introduction of linear corrections or even non-linearities in $\lambda$ constitutes a challenging future perspective. One way to partially avoid the need for an extended power range is the freedom gained when the laser position $x_0(t)$ is no longer fixed but controlled in time: it would allow the required amount of heat to be deposited at each location of the cantilever, bypassing the slow diffusion process to reach the stationary state of the temperature field.

As a final consideration, let us rewind to our initial motivation: get closer or beat the time scales of the mechanical system. With an acceleration of a factor 30, the ratio of the mechanical and thermal time writes $T_0/t_f = \SI{2}{\mu m}/H$. For our $H=\SI{5}{\mu m}$ thick cantilever, we are thus able to stabilize the temperature field in only $2.5$ oscillations, and we would reach half an oscillation with $H=\SI{1}{\mu m}$. As for the mechanical relaxation time $\tau_r$ which is $Q/\pi$ times longer than the period $T_0$, with a quality factor of the order of $Q=3000$ in our experiment in vacuum, we actually reach thermal steady state in $\tau_r/400$. Such an acceleration even allows stabilizing temperature faster than the ``equilibration'' of the first 12 mechanical resonant modes of the cantilever. Our strategy to accelerate the heat diffusion thus achieves its goal, and could be useful in numerous applications once we reach a meaningful temperature step amplitude.

\medskip

The data that support the findings of this study are openly available in Zenodo~\cite{Pottier-2023-FastHeating-DataSet}.

\acknowledgments 
This work has been financially supported by the Agence Nationale de la Recherche through grant ANR-18-CE30-0013. C.A.P. acknowledges financial support from Grant No.~PID2021-122588NB-I00 funded by MCIN/AEI/ 10.13039/501100011033/ and by ``ERDF A way of making Europe'', as well as funding from European Union for the support from Horizon Europe - Marie Sklodowska-Curie 2021 programme through the Posdoctoral Fellowship Ref. 101065902 (ORION). We thank Artyom Petrosyan, Salambô Dago, Ivan Palaia and Sergio Ciliberto for enlightening scientific and technical discussions.

\vfill

\pagebreak

\appendix
\section*{Appendix}

\section{Coefficients \texorpdfstring{$\gamma_n$}{} of the polynomial function \texorpdfstring{$F(t)$}{F(t)}} \label{appendix:gamma}

\begin{figure}[htb]
 \centering
 \includegraphics{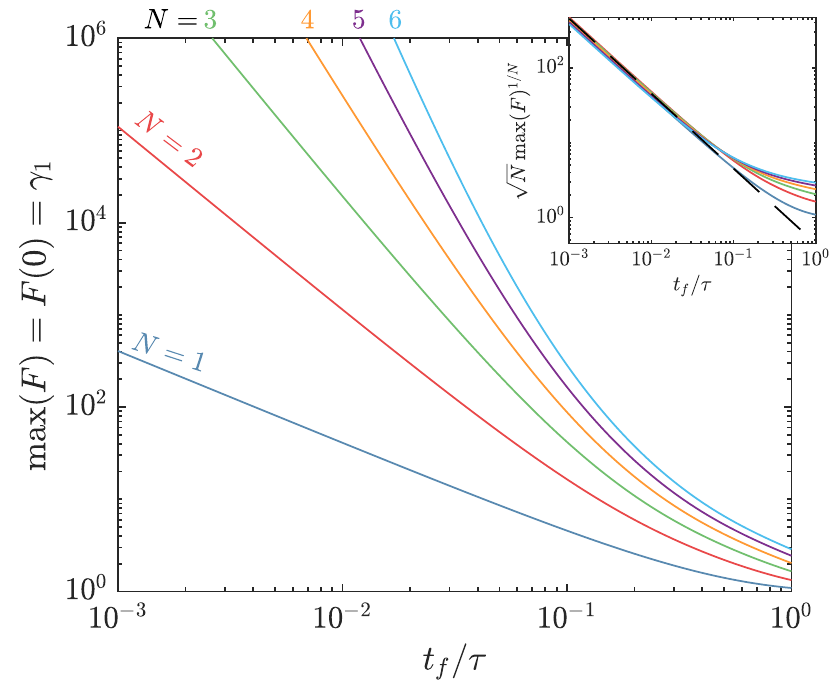} \\ \includegraphics{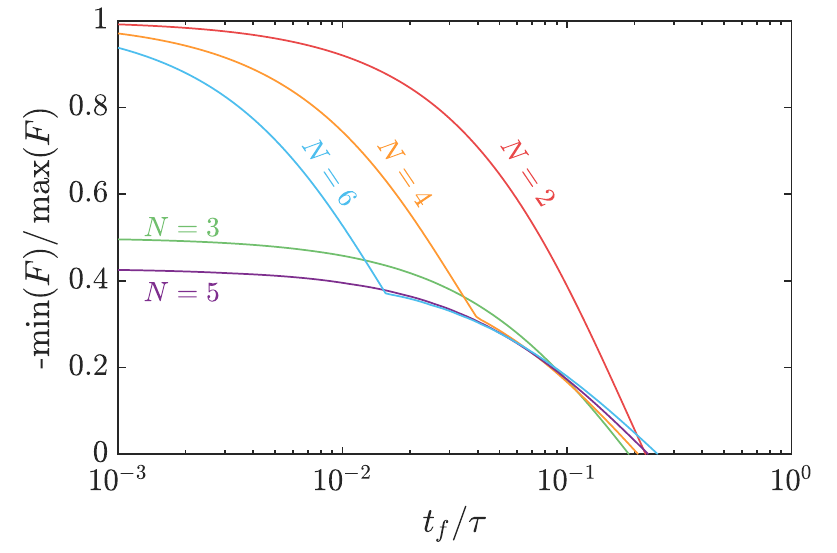}
\caption{\label{fig:maxF} (top) Maximum value of $F(t)$, corresponding to $F(0)=\gamma_1$, for $N=1$ to $6$ and $t_f/\tau=\num{e-3}$ to $1$. Large $N$ and short $t_f$ lead to a huge range of power needed during the transient. A scaling $\max(F)\sim[0.45\tau/(t_f\sqrt{N})]^{N}$ is observed, as illustrated in the inset (dashed line is $0.45\tau/t_f$). (bottom) Minimum value of $F(t)$, normalized by its maximum value, for $N>1$ (for $N=1$, $F(t)>0$ at all times - no cooling is necessary). The cooling power is always smaller in absolute value that the heating one.}
\end{figure}

Eq.~\ref{eq:systemGamCoef} can be written as:
\begin{align}
    \sum_{m=1}^{N} A_{m,n}\gamma_m = 1,\ \mathrm{for}\ n=1\ \mathrm{to}\ N, \\
\mathrm{with} \ \
    A_{m,n} = \int_0^1s^{m-1}k_n^2\tilde{t}_fe^{k_n^2\tilde{t}_f(s-1)}ds.
\end{align}
Using integrations by parts, we compute the coefficient $A_{m,n}$ analytically with the following recurrence relation:
\begin{align}
    A_{1,n} &= 1-e^{-k_n^2\tilde{t}_f} \\
    A_{m+1,n} &= 1-\frac{m}{k_n^2\tilde{t}_f}A_{m,n}
\end{align}
The $N\times N$ square matrix $A$ defined by its coefficients $A_{m,n}$ can then be inverted using a symbolic math solver, and the coefficients $\gamma_m$, solutions of $A [\gamma_m] = [1]$, computed analytically by $[\gamma_m] = A^{-1}[1]$. This approach avoids any rounding errors when computing the integrals in Eq.~\ref{eq:systemGamCoef} or singularities when inverting the matrix $A$. Some examples of coefficients $\gamma_m$ are given in Table \ref{table:gamma} for $\tau/t_f=10$ to $100$ and $N=1$ to $4$, while Fig.~\ref{fig:maxF} reports the maximum and minimum values of $F(t)$ for $\tau/t_f=1$ to $1000$ and $N=1$ to $6$.

\begin{table}[!b]
\center
\begin{tabular}{|cc|r|r|r|r|}
 \hline
  & \hspace{2mm}\hspace{2mm} & \hspace{2mm}$N=1$ ~& \hspace{2mm}$N=2$ ~& \hspace{2mm}$N=3$ ~& \hspace{2mm}$N=4$~\\
 \hline 
 \hline 
 \multirow{4}{5em}{$\tau/t_f=10$}& $\gamma_1$ & 4.57 ~& 16.5 ~& 42 ~& 88.1~\\
 &$\gamma_2$ & & -22.9 ~& -155 ~& -585~\\
 &$\gamma_3$ & & & 122 ~& 1040~\\
 &$\gamma_4$ & & & & -548~\\
 \hline
 \hline
 \multirow{4}{5em}{$\tau/t_f=30$}& $\gamma_1$ & 12.7 ~& 113 ~& 641 ~& 2749~\\
 &$\gamma_2$ & & -198 ~& -3204 ~& -26156~\\
 &$\gamma_3$ & & & 2926 ~& 57017~\\
 &$\gamma_4$ & & & & -34647~\\
 \hline
 \hline
 \multirow{4}{5em}{$\tau/t_f=100$}& $\gamma_1$ & 41 ~& 1141 ~&  19361 ~& 238632~\\
 &$\gamma_2$ & & 2191 ~& ~-109748 ~& -2662356~\\
 &$\gamma_3$ & & & 106680 ~& 6372575~\\
 &$\gamma_4$ & & & & ~-4126391~\\
 \hline
\end{tabular}
\caption{\label{table:gamma}Coefficients $\gamma_n$ of the polynomial function $F(t)$ found solving Eq.~\eqref{eq:systemGamCoef} that allow to cancel successively the four first thermal modes for a protocol duration $t_f$ such that $\tau/t_f=10$, $\tau/t_f=30$ and $\tau/t_f=100$.}
\end{table}

As physically expected, shorter values for $t_f$ or larger number $N$ of modes canceled involve much greater costs. The maximum value of $F(t)$, always occurring at $t=0$, behaves as $(\sqrt{N}t_f/\tau)^{-N}$. This scaling can be used to extrapolate the very large power requirements at strong accelerations and large mode numbers. From the theoretical point of view, no pathologies of any kind are found for the values obtained. However, from the practical point of view, very high intensities could lead to implementation problems for a standard experimental setup. Moreover, the assumption of constant thermal conductivity is jeopardized and non-linear effects beyond the simple heat equation considered herein should be taken into account.

\section{Independence of the protocol on the laser location} \label{appendix:x0}

\begin{figure*}[htb]
 \centering
 \includegraphics{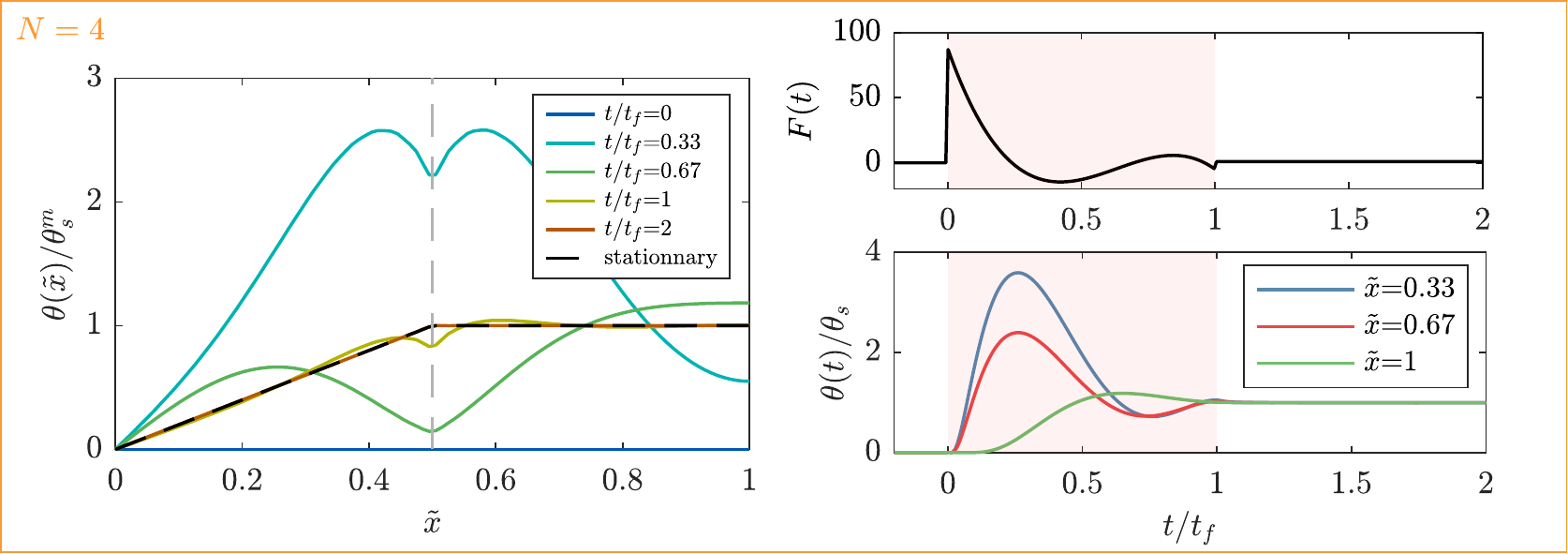}
\caption{\label{fig:x0} Numerical simulation of the protocol for the same parameters as in the bottom panel of Fig.~\ref{fig:AllProtocol}: $N=4$, $t_f=0.1\tau$, but a different value of the heating location, $\Tilde{x}_0=0.5$.}
\end{figure*}

In Fig.~\ref{fig:x0}, we present a numerical check of the validity of our acceleration protocol for a different value of the laser location. More specifically, we take the same parameters as in the bottom panel of Fig.~\ref{fig:AllProtocol}: $N=4$, $t_f=0.1\tau$, but $\Tilde{x}_0=0.5$.

\section{Calibration of the sensitivity coefficient \texorpdfstring{$\beta$}{}} \label{appendix:beta}

\begin{figure}[bht]
 \centering
 \includegraphics{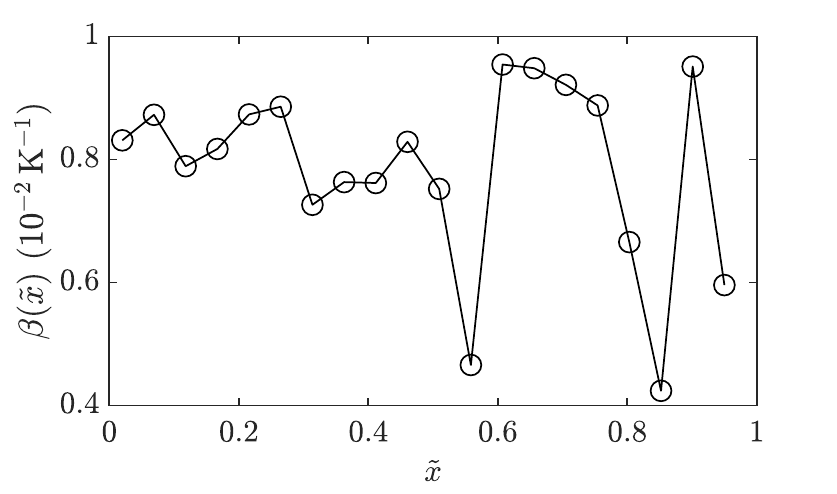}
\caption{\label{fig:sensitivityBeta} Calibration of the sensitivity coefficient $\beta$. This quantity makes it possible to convert the variation of reflectivity into temperature change, $\theta(x)=\beta^{-1}(x)\Delta R(x)/R(x)$. The relatively large range of dispersion for the sensitivity $\beta$ is mainly due to variations of the thickness along the cantilever length.}
\end{figure}

Since the cantilever thickness slightly varies along its length, the sensitivity coefficient $\beta$ depends on the position $x$. To calibrate the function $\beta(x)$, we impose a uniform temperature profile (positioning the heating beam onto the chip) and we measure the induced change in reflectivity $\Delta R(x)$ at each location $x$. At the same time, we measure the shift of the mechanical resonance frequencies by analyzing the thermal noise driven fluctuations of position of the reflected probe beam. This frequency shift leads to a calibrated measurement of the imposed temperature change~\cite{aguilar_sandoval_resonance_2015, Pottier-2021-JAP}. The measured coefficients $\beta$ to convert the reflectivity into temperature at all probed positions $x$ are displayed in Fig.~\ref{fig:sensitivityBeta}. 

\section{Transient temperature during the protocol}
\label{appendix:transient}

\begin{figure}[!b]
 \centering
 \includegraphics{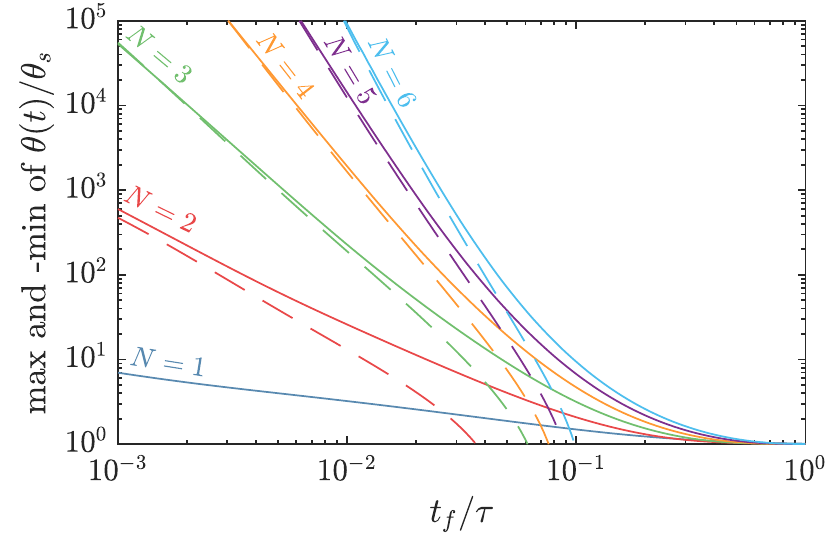}
\caption{\label{fig:Tmax} Maximum (plain line) and opposite of the minimum (dashed line) values of $\theta(t)$ for $N=1$ to $6$, $t_f/\tau=\num{e-3}$ to $1$, and $\Tilde{x}_0=0.95$. Large $N$ and short $t_f$ lead to a huge transient temperature, limiting the magnitude of achievable temperature step $\theta_s$. No negative values of $\theta(t)$ are observed for $N=1$.}
\end{figure}

As illustrated in Figs.~\ref{fig:AllProtocol} and \ref{fig:acce30}, the temperature can present overshoots and undershoots quite far from the target during (and after) the protocol. We plot in Fig.~\ref{fig:Tmax} the range of temperature explored according to the numerical resolution of the heat equation for the drivings defined in Appendix \ref{appendix:gamma}. Transient temperatures can be orders of magnitude larger than the target step, limiting the applicability of very fast acceleration to small $\theta_s$.

\begin{figure}[tbh]
 \centering
 \includegraphics{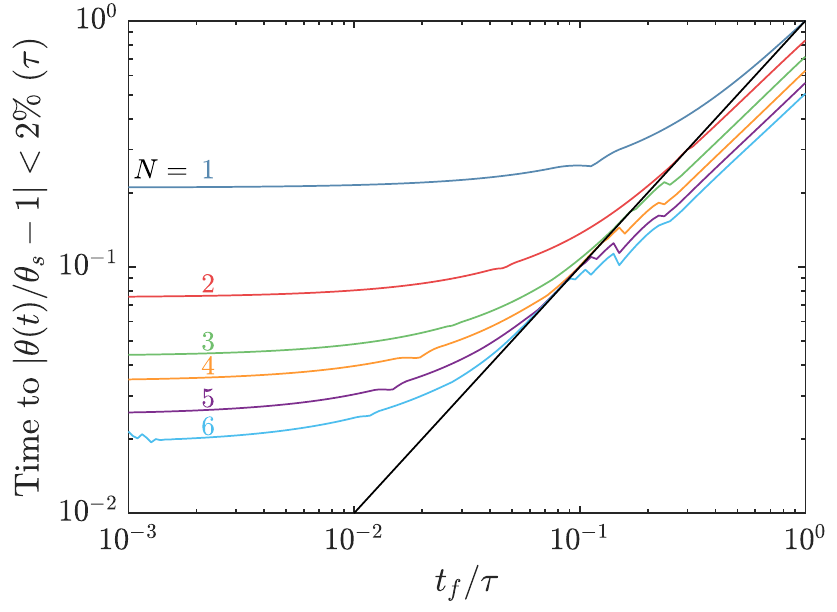}
\caption{\label{fig:t2pc} Time (in units of $\tau$) to reach the target with a 2\% tolerance for $N=1$ to $6$, $t_f/\tau=\num{e-3}$ to $1$, and $\Tilde{x}_0=0.95$. This time is defined as the latest instant when $|\theta(t,x)/\theta_s(x)-1|>2\%$, for $x=0.33$, $0.67$ and $x_0$. $\theta(t,x)$ is computed from numerical resolution of the heat equation for the drivings defined in Appendix \ref{appendix:gamma}. Achieving this goal requires $1.8\tau$ with a power step protocol, and $0.4\tau$ with a perfect feedback loop (temperature step protocol). The black line is the duration of the protocol $t_f/\tau$: curves below it reach the target actually faster than the protocol itself.}
\end{figure}

As a further insight in the dynamics of the temperature and the acceleration provided by our protocols, we report in Fig.~\ref{fig:t2pc} the time to reach the target with a 2\% tolerance. This time depends on the choice of $t_f$ and $N$ obviously, but also on the relaxation of remaining modes after $t_f$. For very large acceleration ($t_f<\tau/100$), the amplitude of the remaining temperature field at $t_f$ is significant and though decaying very fast, further delays the effective reach of the vicinity of the target. Effective accelerations over 100 are thus out of reach in practice.
 
\section{Response to a temperature step (perfect feedback loop)} \label{appendix:tempstep}

We provide here the solution for the heat equation \eqref{eq:heatdiff} without a source term:
\begin{equation}
\label{eq:ap-diff}
\partial_t \theta = D \partial_x^2 \theta,
\end{equation}
with $D=\lambda/(\rho c_p)$ and the following boundary conditions:
\begin{equation}
\theta(0,t)= 0 ,\quad 
\theta(x_0,t)=\theta_s^m , \quad
\left. \partial_x\theta(x,t) \right|_{x=L}=0, 
\end{equation}
with $0<x_0<L$, and initial condition:
\begin{equation}
\theta(x,0)= 0.
\end{equation}
Note that, since there is an extra ``boundary'' condition at $x=x_0$, i.e. within the interval $0\leq x\leq L $, the equation should be solved separately for the two intervals $0 \leq x < x_0$ and $x_0 < x \leq L$. The stationary solution is
\begin{equation}
\frac{\theta_s(x,t)}{\theta_{s}^{m}}=\begin{cases}
\frac{x}{x_{0}} & 0\le x\le x_{0},\\
1 & x_{0}\le x\leq L
\end{cases}
\end{equation}
After expanding $\Delta \theta(x,t) = \theta(x,t) - \theta_s(x)$ in the corresponding eigenbasis, one gets
\begin{widetext}
\begin{equation}
\label{eq:ap-sol}
\frac{\theta(x,t)}{\theta_{s}^{m}}=\begin{cases}
\displaystyle  \frac{x}{x_{0}} + \frac{2}{\pi}\sum_{n=1}^{\infty} \frac{(-1)^n}{n} \sin \left( n \pi \frac{x}{x_0}\right) \exp\left(- \frac{ n^2 \pi^2 D}{x_0^2}t\right) & 0\le x\le x_{0},\\
\displaystyle 1 - \frac{4}{\pi}\sum_{n=1}^{\infty} \frac{1}{2n-1} \sin \left[ \left( n - \frac{1}{2} \right) \pi \frac{x-x_0}{L-x_0} \right] \exp \left[ - \frac{(2n-1)^2 \pi D}{4(L-x_0)^2} t\right]& x_{0}\le x\leq L
\end{cases}
\end{equation}
\end{widetext}
Characteristic relaxation times to the left and to the right of the point $x_0$ are proportional to $\tau_\text{left} = x_0^2 /D$  and   $\tau_\text{right} =( L-x_0)^2/D$ respectively. With respect to the power step solution, assuming $x_0\sim L$, the slowest time constant in the exponentially decaying functions is four time smaller. This acceleration stems from the change of modes from $k_n = (n-1/2) \pi$ to $k_n= n \pi$, due to the change of the relevant boundary condition from Neumann to Dirichlet.

Rather than considering Eq.~\eqref{eq:ap-diff} separately in the intervals $0 \leq x < x_0$ and $x_0 < x \leq L$, we could have instead considered Eq.~\eqref{eq:heatdiff} complemented with Eq.~\eqref{eq:sourceterm}, i.e. with a localized $\delta(x-x_0)$ forcing term proportional to $P(t)$. Thus, the perfect feedback loop introduced in this appendix can be experimentally implemented with a feedforward protocol with the laser power

\begin{equation}
P(t)=  \frac{\lambda S \theta_s^m}{a} \left[ \frac{\vartheta_3 (e^{-\frac{D \pi^2 t }{x_0^2}})}{x_0} + \frac{\vartheta_2(e^{-\frac{D \pi^2 t}{(L-x_0)^2}})}{L-x_0} \right].
\end{equation}
This latter relation is obtained from direct integration of the heat equation \eqref{eq:heatdiff} over an infinitesimal interval around $x_0$, substituting for  $\theta(x,t)$  the solution \eqref{eq:ap-sol}. Above, $\vartheta_a(q)$, with $a\in\{2,3\}$ stands for the elliptic theta function $\Theta_a(u,q)$ evaluated in $u=0$: $\vartheta_a(q)=\Theta_a(0,q)$, with
\begin{align}
\Theta_2(u,q)&=2q^{1/4} \sum_{n=0}^{\infty} q^{n(n+1)} \cos[(2n+1 )u],\\
\Theta_3(u,q)&=1+2 \sum_{n=1}^{\infty} q^{n^2} \cos(2nu).    
\end{align}

\bibliography{FastHeating}

\begin{thebibliography}{52}
\bibitem [{sup()}]{suppmat-movies}%
  \BibitemOpen
  \href@noop {} {}\bibinfo {note} {Movies supplied as ancillary files: {\tt
  tf=0.1tau-N=X.mov}, with {\tt X=1} to {\tt4}, compare the time evolution of
  the temperature profile for a power step, a temperature step (perfect
  feedback loop), and our acceleration protocols for $N=1$ to $4$.}\BibitemShut
  {Stop}%
\end{thebibliography}

\end{document}